\documentclass[prd,aps,tightenlines,a4paper,12pt]{revtex4}

\usepackage{graphicx}
\usepackage{bm}
\usepackage{url}

\newcommand{\ve}[1]{\mbox{\boldmath$#1$}}

\def\source{{\rm 0}}
\def\obs{{\rm 1}}

\usepackage{bm}
\usepackage{graphicx}
\usepackage{hyperref}
\usepackage{amsmath}
\usepackage{verbatim}
\usepackage{epsf}
\usepackage{wasysym}

\begin{document}

\title{Light-propagation in the gravitational field of moving quadrupoles}
\author{Sven \surname{Zschocke}}
\affiliation{
Lohrmann Observatory, Dresden Technical University,\\
Mommsen Str. 13, D-01062 Dresden, Germany\\
}

\begin{abstract}
\begin{center}
{\it GAIA-CA-TN-LO-SZ-007-1}

\medskip

\today

\end{center}
A simplified formula for light-deflection in the quadrupole field of moving massive bodies has been obtained in 
\cite{Report_Zschocke_Klioner_1,Report_Zschocke_Klioner_2,Article_Zschocke_Klioner}, which will be applied for Gaia 
data reduction. So far, in Gaia data reduction it has been assumed that the positions of the giant planets should be 
computed at the retarded instant of time. The problem of light-deflection due to quadrupole field of moving planets 
has been re-considered in \cite{Kopeikin_Makarov}. According to their solution, 
the position and velocity of the massive body have to be taken at retarded time. We show that the solution 
given in \cite{Kopeikin_Makarov} coincides with our simplified quadrupole formula obtained 
in \cite{Report_Zschocke_Klioner_1,Report_Zschocke_Klioner_2,Article_Zschocke_Klioner}. This coincidence implies 
that the positions of giant planets have, in fact, to be taken at retarded time. 
\end{abstract}

\maketitle

\newpage

\tableofcontents

\newpage

\section{Introduction}\label{Introduction}

An essential experiment of Gaia mission to test relativity concerns the new effect of light-deflection at giant planets
due to their quadrupole field. Analytical solutions of light deflection by a quadrupole field of a massive body  
are well-known and have been investigated by many authors 
\cite{Ivanitskaya,Epstein_Shapiro1,Richter_Matzner,Cowling1,Klioner1,Klioner_Kopejkin,Klioner2,Klioner_Blankenburg}.
Because these formulas are rather complicated, they imply massive computations of quadrupole light deflection and are 
too time-consuming for Gaia data reduction. Therefore, we have derived a simplified quadrupole formula in 
\cite{Report_Zschocke_Klioner_1,Report_Zschocke_Klioner_2,Article_Zschocke_Klioner}, which is suitable for a 
time-efficient computation of quadrupole light-deflection on microarcsecond level of accuracy.

In reality, while the light-signal is being emitted at a position $\ve{x}_{\source}$ at time moment $t_0$ and 
received at position $\ve{x}_{\obs}$ at a time moment $t_1$, the massive body chances the position 
from $\ve{x}_A (t_0)$ to $\ve{x}_A (t_1)$. Therefore, it is not obvious at which coordinate time $t_0 \le t \le t_1$ 
the position of the massive body $\ve{x}_A (t)$ has to be chosen in the formula of quadrupole light-deflection.
Up to now, in Gaia data reduction it is tacitly and implicitly assumed that the positions of multipoles 
(Jupiter and Saturn) should be computed at the retarded moments of time, given by the implicit light-cone equation:
\begin{eqnarray} s_1^A = t_1 - \frac{\left| \ve{x}_1 - \ve{x}_A \left(s_1^A\right) \right|}{c}\,,
\label{retarded_time_1} 
\end{eqnarray} 

\noindent
where $\ve{x}_1$ is the position of observer at observation time $t_1$ and $\ve{x}_A \left(s_1^A\right)$ is the
coordinate of massive body at retarded time moment $s_1^A$. However, no theoretical proof of this assumption 
has been given. This problem has recently been solved by \cite{Kopeikin_Makarov}. In our report we show that 
the solutions in \cite{Report_Zschocke_Klioner_1,Report_Zschocke_Klioner_2,Article_Zschocke_Klioner} and 
in \cite{Kopeikin_Makarov} agree with each other, implying that in the quadrupole formula the position 
of massive body has indeed to be taken at the retarded instant of time.

\section{Simplified quadrupole formula}\label{Section0}

In \cite{Report_Zschocke_Klioner_1,Report_Zschocke_Klioner_2,Article_Zschocke_Klioner} a simplified quadrupole formula 
has been derived which takes into account only those terms relevant for microarcsecond level of accuracy. 
In this Section we will give the main steps and results. 
Consider a gravitational field of $N$ massive bodies $A$, and the positions of these individual massive bodies 
are $\ve{x}_A$. A light-ray is being emitted at a position $\ve{x}_{\source}$ at time moment $t_0$ and received at position
$\ve{x}_{\obs}$ at a time moment $t_1$. The unit coordinate direction of the light propagation at the moment of 
observation reads $\displaystyle \ve{n} = \frac{\dot{\ve{x}} (t_{\obs})}{\left|\dot{\ve{x}}(t_{\obs})\right|}$ and the 
unit tangent vector of light path at infinitely past is 
$\displaystyle \ve{\sigma} = \lim_{t \rightarrow - \infty} \,\frac{\dot{\ve{x}} (t)}{c}$. 
Then, in post-Newtonian order, the transformation $\ve{\sigma}$ to $\ve{n}$ is given by \cite{Klioner1,Klioner2}
\begin{eqnarray}
\label{lightpath_B}
\ve{n} &=& \ve{\sigma} + \sum \limits_i \delta \ve{\sigma}_i+{\cal O} \left(c^{-4}\right),
\end{eqnarray}

\noindent
where the sum runs over individual terms of various physical origin, that means monopole gravitational field, quadrupole 
field and higher multipole fields. The spherical symmetric part (monopole 
field) due to one massive body $A$ is given, for instance, by Eq.~(102) in \cite{Article_Klioner_Zschocke}.

Here, we are only interested at the quadrupole light deflection term.
For one massive body $A$, the quadrupole light-deflection is given as follows 
(see Eq.~(40) in \cite{Report_Zschocke_Klioner_2} or Eq.~(9) in \cite{Article_Zschocke_Klioner}):
\begin{eqnarray}
\delta \ve{\sigma}_{\rm Q} &=& - \frac{G\,M_A}{c^2}\,J_2^A\,\frac{P_A^2}{d_A^3}\,
\left(2 + 3\,\cos \psi - \cos^3 \psi\right)
\nonumber\\
\nonumber\\
&& \times \left[\left(1 - \left(\ve{\sigma}\cdot\ve{e}_3\right)^2 - 4\,\left(\ve{n}_A\cdot\ve{e}_3 \right)^2 \right) 
\ve{n}_A + 2\left(\ve{n}_A\cdot\ve{e}_3\right) \ve{e}_3
- 2\left(\ve{\sigma}\cdot\ve{e}_3 \right) \left(\ve{n}_A\cdot\ve{e}_3\right)\ve{\sigma}\right]\,.
\label{Eq_20_B}
\end{eqnarray}

\noindent
Here, $M_A$ is the mass of body $A$, $c$ is the speed of light, $G$ is the gravitational constant, and the impact vector
\begin{eqnarray}
\ve{d}_A &=& \ve{\sigma} \times \left(\ve{r}_{\obs}^A \times \ve{\sigma}\right),
\label{Eq_15}
\end{eqnarray}

\noindent
where $\ve{r}_{\obs}^A = \ve{x}_{\obs} - \ve{x}_A$, and the absolute value $d_A = \left|\ve{d}_A\right|$.
Furthermore, the angle $\psi = \delta \left(\ve{\sigma},\ve{r}_{1}^A\right)$, the unit vector along the axis of symmetry 
(rotational axis of massive body) is denoted by $\ve{e}_3$, $P_A$ denotes the equatorial radius,  
and $J_2^A$ is the coefficient of second zonal harmonic of the gravitational field of massive body $A$.
The unit vector $\ve{n}_A$ is defined by 
\begin{eqnarray}
\ve{n}_A &=& \frac{\ve{d}_A}{d_A}\,.
\label{Eq_11}
\end{eqnarray}

\noindent
Since the massive bodies move, the question arises at which instant of time the positions of the massive bodies 
have to be chosen. This problem has been solved in \cite{Kopeikin_Makarov}, and the main results 
of this work will be subject of the next Section.

\section{Quadrupole formula by Kopeikin $\&$ Makarov}\label{Section1}

\subsection{Description of the approach}

The light-deflection at moving monopoles and quadrupoles has been re-investigated by in \cite{Kopeikin_Makarov}. 
In this Subsection we describe the basic steps of this approach. The gravitational field is described by 
$g_{\alpha\beta} = \eta_{\alpha\beta} + h_{\alpha\beta}$ where $h_{\alpha \beta}$ is the metric perturbation in 
post-Minkowski apprximation. Using harmonic gauge, the linearized Einstein equations for the pertubation 
$h_{\alpha \beta}$ are homogeneus wave equations,
\begin{eqnarray}
\left(- \frac{\partial^2}{\partial t^2} + {\ve{\nabla}}^2\right) h_{\alpha \beta} &=& 0\,.
\label{Eq_KM_10}
\end{eqnarray}

\noindent
A general solution of (\ref{Eq_KM_10}) is given in terms of multipole expansion 
\cite{Blanchet_Damour,Thorne}, where the terms in the 
perturbation $h_{\alpha \beta}$ are taken at the retarded instant of time $s^A_1$. The retardation follows directly 
from the retarded (causal) solution of the homogeneus wave equations (\ref{Eq_KM_10}). 

In the approach \cite{Kopeikin_Makarov} the monopole, dipole and quadrupole terms of the general multipole expansion of 
$h_{\alpha \beta}$, given in Eqs.~(12) - (14) in \cite{Kopeikin_Makarov}, are taken into account. 
The center of coordinate system is shifted from the mass center of the massive body by a spatial distance. 
Then, by means of parallel axis theorem (Huygens-Steiner theorem), they apply general expressions for the 
quadrupole moment and their time derivative, see Eq.~(17) and Eq.~(20) in \cite{Kopeikin_Makarov}.

Then, in \cite{Kopeikin_Makarov} the geodesic equation is rewritten into a considerably simplier form, 
given by Eq.~(19) in \cite{Kopeikin_Schafer}. The integration 
leads to the expressions (28) and (29) in \cite{Kopeikin_Makarov}. Using their developed integration method, 
presented in part by Eqs.~(30) and (31) in \cite{Kopeikin_Makarov}, they succeed to integrate analytically 
the differential equation (29) in \cite{Kopeikin_Makarov}. Furthermore, in \cite{Kopeikin_Makarov} all those 
terms are neglected which they proof to contribute less than $1$ microarcsecond. 
The result of their approach is finally given by Eqs.~(39) - (41) in \cite{Kopeikin_Makarov}.

\subsection{The analytical solution for quadrupole light-deflection}

In Eq.~(44) in \cite{Kopeikin_Makarov}, the following form for 
the light-deflecetion of sources at infinite distances due to one moving quadrupole $A$ has been given: 
\begin{eqnarray}
\delta \ve{\sigma}_{\rm Q}^A &=& 4\left(1-\frac{\ve{\sigma}\cdot\ve{v}_A}{c}\right) 
\frac{G\,M_A}{c^2}\,J_2^A\,\frac{P_A^2}{d_A^3}
\nonumber\\ 
\nonumber\\ 
&& \times \left[\left[\left(\ve{e}_3 \cdot\ve{n}_A\right)^2 - \left(\ve{e}_3\cdot\ve{m}_A\right)^2\right]\,\ve{n}_A 
- 2 \left(\ve{e}_3\cdot\ve{n}_A\right) \left(\ve{e}_3\cdot\ve{m}_A\right) \ve{m}_A\right]\,. 
\label{Eq_5}
\end{eqnarray}

\noindent
It is essential to noticed, that all time-dependent quantities in Eq.~(\ref{Eq_5}), that is
$\ve{x}_A$, $\ve{v}_A$ and $\ve{e}_3$, and therefore also $\ve{n}_A$ and $\ve{m}_A$, are computed at retarded
instant of time $s_1^A$ given by Eq.~(\ref{retarded_time_1}). It should be noticed that in Eq.~(\ref{Eq_5}) we have 
omitted a term proportional to the displacement of the planetary center from the origin of the 
coordinate system which is an artificial term of the approach and not relevant for Gaia data reduction. 
Furthermore, the unit vector $\ve{m}_A$ is defined by 
\begin{eqnarray}
\ve{m}_A = \ve{\sigma} \times \ve{n}_A\,.
\label{Eq_10}
\end{eqnarray}

\noindent
In Eq.~(\ref{Eq_5}) we have used our notational conventions in \cite{Report_Zschocke_Klioner_2,Article_Zschocke_Klioner}.
As mentioned in the introductionary Section, we have replaced the vector $\ve{k}$ by $\ve{\sigma}$, since 
$\ve{\sigma} = \ve{k} + {\cal O} \left(m\right)$, 
that means such a replacement would cause effects of higher order beyond the post-Newtonian approximation.

\subsection{Comparison of the formula of quadrupole light-deflection}

The quadrupole light-deflection (\ref{Eq_5}) can be estimated by
\begin{eqnarray}
\left|\delta \ve{\sigma}_{\rm Q}^A \right| &\le& 4\,\left(1-\frac{\ve{\sigma}\cdot\ve{v}_A}{c}\right)
\frac{G\,M_A}{c^2}\,\frac{\left|J_2^A\right|}{P_A}\,.
\label{estimate_1}
\end{eqnarray}

\noindent
Accordingly, for Jupiter and Saturn we have 
\begin{eqnarray}
\left|\ve{\sigma}_{\rm Q}^{\rm Jupiter}\right| &\le& 
\left(1-\frac{\ve{\sigma}\cdot\ve{v}_{\rm Jupiter}}{c}\right) 240\,\mu{\rm as}\,,
\label{quadrupole_jupiter}
\\
\nonumber\\
\left|\ve{\sigma}_{\rm Q}^{\rm Saturn}\right| &\le& 
\left(1-\frac{\ve{\sigma}\cdot\ve{v}_{\rm Saturn}}{c}\right) 95\,\mu{\rm as}\,,
\label{quadrupole_saturn}
\end{eqnarray}

\noindent
while for other massive bodies of the solar system we find considerably smaller values. Furthermore, the
orbital speed of these planets in respect to the barycenter is of the order $v_A \sim 10^{-4}\,c$. Hence, the
contribution of the velocity-term in the quadrupole light-deflection is by far less than about $0.1\,\mu{\rm as}$
and can be neglected for Gaia astrometric accuracy. Thus, the quadrupole light-deflection formula (\ref{Eq_5}) can be 
simplified as follows:
\begin{eqnarray}
\delta \ve{\sigma}_{\rm Q}^A &=& 4 \,\frac{G\,M_A}{c^2}\,J_2^A\,\frac{P_A^2}{d_A^3}
\nonumber\\
\nonumber\\
&& \times \left[\left[\left(\ve{e}_3\cdot\ve{n}_A\right)^2 - \left(\ve{e}_3\cdot\ve{m}_A\right)^2\right]\,\ve{n}_A
- 2 \left(\ve{e}_3\cdot\ve{n}_A\right) \left(\ve{e}_3\cdot\ve{m}_A\right) \ve{m}_A\right]\,.
\label{Eq_20_A}
\end{eqnarray}

\noindent
Eq.~(\ref{Eq_20_A}) determines the quadrupole light-deflection for moving massive bodies, while the 
rotational axis $\ve{e}_3$ and the unit vectors $\ve{n}_A$ and $\ve{m}_A$ have to be computed at retarded
instant of time $s_1^A$ given by Eq.~(\ref{retarded_time_1}). 

The expression given by Eq.~(\ref{Eq_20_A}) in \cite{Kopeikin_Makarov} coincides with our expression given 
in Eq.~(\ref{Eq_20_B}), see \cite{Report_Zschocke_Klioner_1,Report_Zschocke_Klioner_2,Article_Zschocke_Klioner}. 
This can be shown by means of the relation 
$\cos^2 x = 1 - \sin^2 x$, that is $\left(\ve{e}_3\cdot\ve{m}_A\right)^2 = 1 - \left(\ve{e}_3\times \ve{m}_A\right)^2$.
Then, by taking into account $\ve{e}_3\times \ve{m}_A = \ve{e}_3\times \left(\ve{\sigma}\times \ve{n}_A\right)$
we obtain the relation
\begin{eqnarray}
\left(\ve{e}_3\cdot\ve{m}_A\right)^2 &=& 1 - \left(\ve{e}_3\cdot\ve{n}_A\right)^2 
- \left(\ve{e}_3\cdot\ve{\sigma}\right)^2\,.
\label{Eq_30}
\end{eqnarray}

\noindent
Inserting (\ref{Eq_30}) into (\ref{Eq_20_A}) we obtain
\begin{eqnarray}
\delta \ve{\sigma}_{\rm Q} &=& - 4 \,\frac{G\,M_A}{c^2}\,J_2^A\,\frac{P_A^2}{d_A^3}
\nonumber\\
\nonumber\\
&& \times \left[
\left[1 - 2 \left(\ve{e}_3\cdot\ve{n}_A\right)^2 - \left(\ve{e}_3\cdot\ve{\sigma}\right)^2\right]\,\ve{n}_A
+ 2 \left(\ve{e}_3\cdot\ve{n}_A\right) \left(\ve{e}_3\cdot\ve{m}_A\right) \ve{m}_A\right]\,.
\label{Eq_20_C}
\end{eqnarray}

\noindent
Furthermore, for the vector $\ve{e}_3$ we have to use the linear combination in terms of the unit vectors 
$\ve{d_A}$, $\ve{\sigma}$ and $\ve{m}_A$, that means $\ve{e}_3 = \left(\ve{e}_3\cdot\ve{\sigma}\right)\ve{\sigma} 
+ \left(\ve{e}_3\cdot\ve{n}_A\right)\ve{n}_A + \left(\ve{e}_3\cdot\ve{m}_A\right)\ve{m}_A$, from which we conclude 
\begin{eqnarray}
\left(\ve{e}_3\cdot\ve{m}_A\right)\ve{m}_A &=& \ve{e}_3 
- \left(\ve{e}_3\cdot\ve{\sigma}\right)\ve{\sigma} - \left(\ve{e}_3\cdot\ve{n}_A\right)\ve{n}_A\,.
\label{linear_combination}
\end{eqnarray}

\noindent
Inserting (\ref{linear_combination}) into (\ref{Eq_20_C}), we obtain
\begin{eqnarray}
\delta \ve{\sigma}_{\rm Q}^A &=& - 4\,\frac{G\,M_A}{c^2}\,J_2^A\,\frac{P_A^2}{d_A^3}
\nonumber\\
\nonumber\\
&& \hspace{-0.5cm} 
\times \left[\left(1 - \left(\ve{\sigma}\cdot\ve{e}_3\right)^2 - 4\,\left(\ve{n}_A\cdot\ve{e}_3 \right)^2 \right) 
\ve{n}_A 
+ 2\left(\ve{n}_A\cdot\ve{e}_3\right) \ve{e}_3 - 2\left(\ve{\sigma}\cdot\ve{e}_3 \right) 
\left(\ve{n}_A\cdot\ve{e}_3\right)\ve{\sigma}
\right]\,.
\label{Eq_20_D}
\end{eqnarray}

\noindent
This expression coincides with our quadrupole light-deflection for a massive body, given 
by Eq.~(\ref{Eq_20_B}) if we approximate $2 + 3\,\cos \psi - \cos^3 \psi$ by a factor $4$. 
This approximation follows from $\displaystyle \sin \psi = \frac{d_A}{r_1^A}$, that means  
$\displaystyle 2 + 3\,\cos \psi - \cos^3 \psi = 4 + {\cal O} \left(\frac{d_A^4}{{r_1^A}^4}\right)$.
Thus, we have shown the coincidence of our quadrupole formula given by Eq.~(\ref{Eq_20_B}) 
(that means Eq.~(40) in \cite{Report_Zschocke_Klioner_2} or Eq.~(9) in \cite{Article_Zschocke_Klioner}) with the 
quadrupole formula in \cite{Kopeikin_Makarov} given by Eq.~(\ref{Eq_20_A}). This coincidence implies 
that for Gaia data reduction the position of giant planets has, in fact, to be taken at the retarded instant of time, 
given by the implicit light-cone equation (\ref{retarded_time_1}).

\section{Summary}\label{Summary}

A simplified quadrupole formula for massive bodies has been derived in 
\cite{Report_Zschocke_Klioner_1,Report_Zschocke_Klioner_2,Article_Zschocke_Klioner}, 
which will be used for a time-efficient computation of quadrupole light-deflection in Gaia data reduction.
However, since the massive bodies move during the light-signal travelling, it is not obvious at which coordinate time 
the position of the massive body $\ve{x}_A (t)$ has to be taken in this formula of quadrupole light-deflection.
So far, in Gaia data reduction it has been tacitly assumed that the positions of the giant planets should be computed at 
the retarded instant of time $\ve{x}_A (s_1^A)$. However, no theoretical proof of this assumption has been 
performed. 

Therefore, in this report, we have re-considered the results of \cite{Kopeikin_Makarov} where the light-deflection at 
moving quadrupoles has been determined. We have shown that the quadrupole formula obtained in \cite{Kopeikin_Makarov} 
given by Eq.~(\ref{Eq_20_A}) coincides with our simplified quadrupole formula obtained in 
\cite{Report_Zschocke_Klioner_1,Report_Zschocke_Klioner_2,Article_Zschocke_Klioner} and given by Eq.~(\ref{Eq_20_B}). 
This coincidence implies that the position of massive body has, in fact, to be taken at the retarded instant of time.

\section*{Acknowledgements}

This work was partially supported by the BMWi grants 50\,QG\,0601 and 50\,QG\,0901 
awarded by the Deutsche Zentrum f\"ur Luft- und Raumfahrt e.V. (DLR). 
Enlighting discussions with Professor Sergei A. Klioner are gratefully 
acknowledged.

\end{document}